\documentclass[aps,pre,twocolumn,showpacs,preprintnumbers,amsmath,amssymb]{
revtex4}

\usepackage{graphicx}
\usepackage{dcolumn}
\usepackage{bm}

\begin{document}

\title{Visually-salient contour detection using a V1 neural model with
horizontal connections}
\author{P.\ N.\ Loxley}
\author{L.~M.~Bettencourt}
\affiliation{Center for Nonlinear Studies and T-5, Theoretical Division, Los
Alamos National Laboratory, Los Alamos, New Mexico 87545, USA}
\date{\today}

\begin{abstract}
A convolution model which accounts for neural activity dynamics in the primary
visual cortex is derived and used to detect visually salient contours in images. Image inputs to the model are modulated by long-range horizontal connections, allowing contextual effects in the image to determine visual saliency, i.e.~line segments arranged in a closed contour elicit a larger neural response than line segments forming background clutter. The model is tested on 3 types of contour, including a line, a circular closed contour, and a non-circular closed contour. Using a modified association field to describe horizontal connections the model is found to perform well for different parameter values. For each type of contour a different facilitation mechanism is found. Operating as a feed-forward network, the model assigns saliency by increasing the neural activity of line segments facilitated by the horizontal connections. Alternatively, operating as a feedback network, the model can achieve further improvement over several iterations through cooperative interactions. This model has no dynamical stability issues, and is suitable for use in biologically-inspired neural networks.
\end{abstract}
\maketitle

\section{Introduction}

Salient features in an image have the defining property that they instantly
attract our visual attention towards their locations without the need to scan
the entire image systematically---they are image features which have a ``pop
out" effect \cite{koch}. In addition, such features are in some sense rare or
unique---they have a low probability of being mistaken for other features in an
image, and often best distinguish one particular object category from another
\cite{walker,moosmann,sharon}. This property may allow the visual system to
rapidly segment and classify an image based on its most distinctive features, or
to devote further visual processing resources to specific points in an image. In
evolutionary terms a rapid response to salient features could be critical for
biological systems in predator-prey situations. In computer vision, saliency has
been shown to improve performance in object and feature recognition tasks
\cite{walker,moosmann}.

Saliency is thought to be assigned early in the visual pathway, and models of
the primary visual cortex (V1) are often used for this purpose
\cite{pettet,yen,li98,li99,li01}. One reason is that saliency is assumed to
involve only bottom-up (image-based) mechanisms which group simple features
together and process a scene rapidly and in parallel over the entire visual
field \cite{koch}. Later stages of visual attention probably involve top-down
(task-dependent) cues \cite{itti}. Long range horizontal connections in V1
provide a likely mechanism for rapid integration of information across the
visual field \cite{gilbert,kapadia,bosking}. The salient objects we shall mostly
be concerned with in this work are closed contours---such as the
boundaries of objects and surfaces in images. Simple cells in V1 respond best to
oriented bars and edges like those detected using a gabor filter \cite{jones},
and a contour is represented as a discrete set of line segments in V1. However, any given image will result in many line segments in V1, and most of these will be unrelated to
any salient contour. Long-range horizontal connections modulate neural activity,
and provide a mechanism for grouping together line segments that form
salient contours. For example, two tangents are said to be co-circular when they are tangent to the same circle \cite{parent}. If horizontal connections facilitate co-circularity then neural activity
will be higher (and the feature more salient) when line segments form a circular contour rather than a cluttered background of line segments with random orientations and positions, for example. 

Experiments on contour detection by the human visual system have shown that contour saliency depends on factors such as contour length, curvature, degree of
fragmentation, whether the contour is open or closed, and the statistical
properties of the background relative to the contour
\cite{kapadia,kovacs,braun}. Line segments which are co-linear (i.e. lie on a
straight line and are orientated in the same direction as the line) are highly
salient, but saliency decreases as the line segments are either separated or
their relative orientations are changed \cite{kapadia}. The saliency of a
contour can be increased either by increasing the contour length or by closing the contour, but decreases with increase in the contour curvature \cite{kovacs}. A final
point is that saliency also depends on the statistical properties of the
background---false contours formed by fortuitously aligned background line
segments will decrease contour saliency \cite{braun}.  Field et.~al.~proposed an association
field for grouping line segments that form contours \cite{field}. We will make use of this in Sec.~\ref{sec3} when choosing a form for horizontal connections.

The purpose of this work is to present a neural model with V1 architecture that
can be used to detect visually salient contours in images. We would also like to determine what level of performance is possible in such tasks using a simple V1 architecture. This architecture
includes functionally specific long-range horizontal connections, local feedback
connections, thresholding, and neural orientation preference. A key motivation is to see how a
biophysically-constrained model of V1 can be used to assign saliency in the
10-20 ms period available under certain viewing conditions \cite{keysers}. We would also like to provide a solid framework for including horizontal connections into certain
classes of biologically-inspired hierarchical neural networks used for object
recognition \cite{serre}. We start with a model of neural activity (spike-rate)
dynamics such as used in Refs.~\cite{li98} and \cite{wilson,bressloff,loxley}
for describing populations of V1 neurons. From these dynamical equations a
convolution model is derived under the assumption that horizontal connections
only modulate neural activity rather than drive it. A convolution model has
the advantage that it avoids the stability problems
found in dynamical neural models \cite{li98,li01}. Operating as a feed-forward network, the convolution model increases the activity of line segments
facilitated by the horizontal connections. Alternatively, operating as a feedback network, it increases the activity of facilitated line segments over several iterations through cooperative interactions.

Previous models of salient contour detection can be divided approximately into
two classes: V1-like neural models with horizontal connections
\cite{pettet,yen,li98}, and non-neural models as typified by
Refs.~\cite{zucker,sha,williams,wang,zhu,ren}. Although the non-neural models are interesting in their own right, here our primary interest is
in biologically-inspired visual processing, and we will not further consider
non-neural models. In Ref.~\cite{pettet}, neural elements were represented
by gabor patches, and a dynamical model with facilitatory horizontal connections
and global inhibition was assumed. Horizontal connections involved fitting a
cubic spline between pairs of neighboring patches, then including penalties for
distance, curvature, and change in curvature---such that co-linear and
co-circular neighbors were facilitated. A region was salient when its steady state
response reached a certain out-lier criterion level separating background noise
from signal. In Ref.~\cite{yen}, neural elements were grouped using a model with
facilitatory horizontal connections and various forms of competitive inhibition.
Horizontal connections were formed from an association field similar to that proposed in
Ref.~\cite{field}, and co-circular neighbors were facilitated. Global inhibition was used
to suppress weakly facilitated elements, and the remaining elements were modeled
as phase oscillators. Salient regions exhibited phase synchrony due to strong
facilitatory interactions. In Ref.~\cite{li98}, each edge segment represented a
pair of neural elements, one excitatory and one inhibitory, which were described
using a dynamical model with horizontal connections. Facilitation came from an association field,
and inhibition was given by so-called \textit{flank inhibition} \cite{li98,li01}. Saliency was
related to the activity level at the steady state. In addition, the nonlinear dynamical
properties of this model caused gaps in contours to be filled in, and activity
to leak out across boundaries \cite{li01}. 

The
structure of this paper is as follows: In Sec.~\ref{sec2} the basic model is
derived and explained. In Sec.~\ref{sec3} the model is used to find salient
contours in example images which include a line, a circular closed contour, and a non-circular closed contour. In Sec.~\ref{sec4}, a summary and a discussion are given.

\section{Model}\label{sec2}

The model we consider describes the processing of visual information in a sequence of stages which reside in early parts of the visual pathway. Visual input from an image 
arrives at V1 simple cells, which respond most
strongly to visual features such as a bar or edge of a particular orientation. Dynamical processes then rapidly redistribute
the neural input among the simple cells via functionally-specific long
range
horizontal connections. Once neural activity reaches equilibrium after
approximately 10 ms it is either
directly output to later stages of the visual pathway, or else iterated several times
through thresholding and
local feedback connections before being output to later stages. The most salient regions in the image correspond to the regions of highest activity in the V1 model output. We now
describe the three stages in greater detail.

\subsection{Simple Cell Classical Receptive fields}

Visual input to the model comes from an image, and is given by the set of intensity values
$I({\bf{r}})$ for each point ${\bf{r}}=(x,y)$ in the image. To construct the
simple cell classical receptive fields, this input is convolved with the gabor
function $g_{\phi}({\bf{r}})$ before thresholding is applied, giving
\begin{equation}
I_{\phi}({\bf{r}})=H\left (\int
g_{\phi}({\bf{r}}-{\bf{r}}^{\prime})I({\bf{r}}^{\prime})d{\bf{r}}^{\prime}\right ),
\label{gabor}
\end{equation} 
where the gabor function is
\begin{eqnarray}
g_{\phi}({\bf{r}})&=&g(R_{\phi}{\bf{r}}),\\
R_{\phi}&=&\left(\begin{array}{cc}
\cos{\phi}&-\sin{\phi}\\
\sin{\phi}&\cos{\phi}\\
\end{array}\label{rot}
\right),
\\ \nonumber \\
g({\bf{r}})&=&\exp{\left(-\frac{x^2+\gamma^2y^2}{2\sigma^2}\right)}\cos{
\left(\frac{2\pi}{\lambda}x+\varphi\right)},\label{endgabor}
\end{eqnarray}
and depends on the parameters $\phi$, $\gamma$, $\sigma$, $\lambda$, and
$\varphi$ \cite{jones}. The integral in Eq.~(\ref{gabor}) is taken over the
whole image, and $H$ is a unit step function satisfying: $H(x)=1$ if
$x\geq \kappa$, and $H(x)=0$ if $x< \kappa$. The key parameter here is $\phi$---the neural orientation
preference---which we have made explicit in Eq.~(\ref{gabor}). All other parameters will be considered as being constant for a particular image.

\subsection{Horizontal Connections and Neural Dynamics}

Simple-cell neural populations will be labeled by their position ${\bf{r}}\in R^{2}$, and
orientation preference $\phi\in(0,\pi)$. The dynamics of simple-cell neural populations is
described by the neural activity $u_{\phi}({\bf{r}},t)$, which corresponds to
the local mean cell-body potential. Given a visual input $I_{\phi}({\bf{r}})$
from the previous stage, neural activity is redistributed according to long
range
orientation-specific horizontal connections
$w_{\phi\phi^{\prime}}({\bf{r}}-{\bf{r}}^{\prime})$, which link cell populations
at
${\bf{r}}$ and $\phi$, with cell populations at ${\bf{r}}^{\prime}$ and
$\phi^{\prime}$.
The equation for neural-activity dynamics has a similar form to that used in
coarse-grained models of V1 \cite{li98,wilson,bressloff,loxley}, and is given by
\begin{eqnarray}
&&\left(\tau\frac{\partial }{\partial t}+1\right)u_{\phi}({\bf{r}},t)\nonumber
\\
&&=\sum_{\phi^{\prime}}\int
w_{\phi\phi^{\prime}}({\bf{r}}-{\bf{r}}^{\prime})S\boldsymbol{(}u_{\phi^{\prime}}({\bf{r
}}^{\prime},t)\boldsymbol{)}d{\bf{r}}^{\prime}+I_{\phi}({\bf{r}}),\label{dyn1}
\end{eqnarray}
where $\tau \simeq 10$ ms is the membrane time constant over which incoming
action potential
spikes are smoothed into pulses at the synapses of a neuron, and $S$ is the
population spiking rate, given by 
\begin{equation}
S\left(u\right)=\frac{1}{1+\exp(-(u-\theta)/\sigma^{\prime})},\label{dyn2}
\end{equation}
which is a sigmoid-shaped function, with spiking-rate threshold $\theta$, and
width $\sigma$. The sum in Eq.~(\ref{dyn1}) is over all neural orientation
preferences---presumably forming a discrete set, and the integral is over all neural population positions.

Solutions of Eq.~(\ref{dyn1}) are guaranteed to converge to a stable fixed point
if an associated Lyapunov function exists \cite{loxleyB}. Alternatively, we can
avoid the dynamical stability problem by setting the time derivative in
Eq.~(\ref{dyn1}) to zero, and working with an equilibrium solution. We also let
$\sigma^{\prime}\rightarrow 0$ in Eq.~(\ref{dyn2}) so that we can replace $S$ in Eq.~(\ref{dyn1}) with a unit step function $H$ satisfying: $H(x)=1$ if
$x\geq \theta$, and $H(x)=0$ if $x< \theta$. The final step is to assume
$I_{\phi}({\bf{r}})\geq \theta$ are the only points where
$u_{\phi}({\bf{r}})\geq \theta$, i.e. neural activity is only ever above
threshold
where the input is above threshold. This means horizontal connections
modulate neural activity rather than drive it \cite{gilbert,kapadia,bosking}. We can now replace
$u_{\phi}({\bf{r}})$ with $I_{\phi}({\bf{r}})$ in the argument of $H$, and
Eq.~(\ref{dyn1}) reduces to
\begin{equation}
u_{\phi}({\bf{r}})=\sum_{\phi^{\prime}}\int
w_{\phi\phi^{\prime}}({\bf{r}}-{\bf{r}}^{\prime})H\boldsymbol{(}I_{\phi^{\prime}}({\bf{r
}}^{\prime})\boldsymbol{)}d{\bf{r}}^{\prime}+I_{\phi}({\bf{r}}).\label{eqlm}
\end{equation}
This equation is our convolution model, giving the facilitation due to horizontal connections. The horizontal connections have the form
\begin{equation}
w_{\phi\phi^{\prime}}({\bf{r}})=M_{\phi\phi^{\prime}}
w(R_{\phi^{\prime}}{\bf{r}}),\label{connectivity}
\end{equation}
where $R_{\phi}$ is the rotation matrix from Eq.~(\ref{rot}),
$M_{\phi\phi^{\prime}}$ is a matrix for the orientation preference specificity of
the horizontal connections, and $w({\bf{r}})$ is a function that depends on
the spatial distribution of the connections.

Processing time is limited in many
visual tasks where
bottom-up saliency is required. The equilibrium given by Eq.~(\ref{eqlm}) takes approximately 10 ms to attain, and neural activity can then be directly output to later parts of the visual pathway. Summing neural activity over all orientation preferences gives the total output as
\begin{equation}
u({\bf{r}})=\sum_{\phi} u_{\phi}({\bf{r}}),\label{totalact}
\end{equation} 
and the regions of largest $u({\bf{r}})$ are the regions which are most salient in the image input. Alternatively, if a longer processing time is available, neural activity can be iterated through thresholding and local feedback connections, as described next.

\subsection{Thresholding and Local Feedback}

This stage of the model involves thresholding $u_{\phi}({\bf{r}})$ from Eq.~(\ref{eqlm}) to select only the regions where it attains its largest values, returning the result to Eq.~(\ref{eqlm}) as $I_{\phi}({\bf{r}})$, then re-calculating $u_{\phi}({\bf{r}})$, and so on. The pooling operation of V1 complex cells has previously been modeled by selecting the maximum from each group of simple cell outputs \cite{serre}. Here, we use inhibitory neural dynamics to perform a global thresholding of simple cell outputs. In a more detailed model, the pooling operation of V1 complex cells could be used to perform a local thresholding. The equilibrium solution for spatially-uniform inhibitory dynamics is given by
\begin{equation}
I^{\prime}_{\phi}({\bf{r}})= H(u_{\phi}({\bf{r}})- u),\label{thresh}
\end{equation}
where $H$ is a unit step function with threshold $\theta$, and $u$ determines the strength of spatially uniform
inhibition. The solution given by Eq.~(\ref{thresh}) is a simple
thresholding step: defining $\Delta\theta\equiv u$, then all
$u_{\phi}({\bf{r}})<\theta+\Delta\theta$ is discarded in Eq.~(\ref{thresh}). This is consistent with selecting regions where $u_{\phi}({\bf{r}})$ attains its largest values.

Next, we include the effect of local feedback connections by replacing $I_{\phi}({\bf{r}})$ in Eq.~(\ref{eqlm}) with $I^{\prime}_{\phi}({\bf{r}})$ from Eq.~(\ref{thresh}), then use it to re-calculate $u_{\phi}({\bf{r}})$. The whole process is repeated as many times as possible. Iterating Eqs.~(\ref{eqlm}) and (\ref{thresh}) will mean any visual input not facilitated by the horizontal
connections will eventually be discarded due to thresholding. The number of possible iterations is determined by the processing time available for the specific visual task being carried out. Assuming local feedback and thresholding act faster than horizontal integration, each iteration will require at least 10 ms---meaning that only a few iterations may be possible in many low level visual tasks involving saliency.

\section{Detecting salient contours in example images}\label{sec3}

The model given by Eqs.~(\ref{gabor})--(\ref{endgabor}) and (\ref{eqlm})--(\ref{thresh}) is now used to find salient contours in the example images shown in Fig.~1. A salient line made from 4 co-linear line segments is shown in Fig.~1(a) and will be treated in Sec.~\ref{sec3.1}, a salient circular closed contour is shown in Fig.~1(b) and will be treated in Sec.~\ref{sec3.2}, while a salient non-circular closed contour is shown in Fig.~1(c) and will be treated in Sec.~\ref{sec3.3}.
\begin{figure}
\includegraphics[width=150pt,bb=247 237 370 631,clip=true]{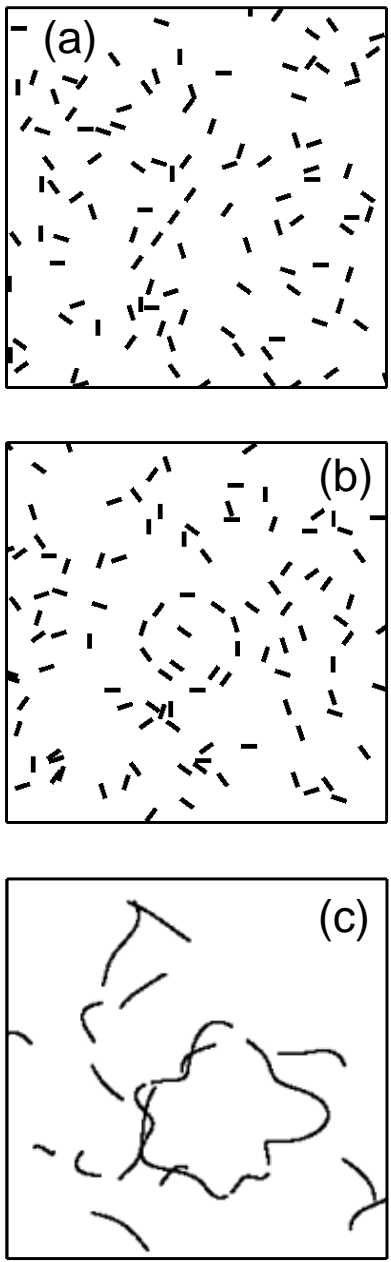}
\caption{Example images of salient contours within cluttered backgrounds. (a) Salient line made from 4 co-linear line segments. (b) Salient circular closed contour. (c) Salient non-circular closed contour.}
\end{figure}

\subsection{Form of Horizontal Connections}

Horizontal connections are parameterized to link neural populations which are neighbors in both ${\bf{r}}$ and $\phi$---i.e.~position and orientation preference. For any two populations with $\phi$ and $\phi^{\prime}$, the elements of $M_{\phi\phi^{\prime}}$ in Eq.~(\ref{connectivity}) are chosen as
\begin{equation}
M_{\phi\phi^{\prime}}=\left\{
\begin{array}{ll}
1,&\mathrm{if}\ |\phi \pmod{\pi}-\phi^{\prime}|\leq\phi_{max},\\
0,&\mathrm{otherwise},
\end{array}\right.
\end{equation}
where mod $\pi$ denotes that a factor of $\pi$ can be added or subtracted from $\phi$ in order to make the inequality true. The parameter $\phi_{max}$ is the size of the $\phi$-neighborhood of neural populations which are connected. The form of $w({\bf{r}})$ used in Eq.~(\ref{connectivity}) is shown in Fig.~2, and is a modified form of the association field proposed in Ref.~\cite{field}. The spatial distribution of connections is anisotropic: strong excitation along the axis of orientation preference (shown as $\phi=0$ in Fig.~2) means that co-linear line segments are strongly facilitated. Crude facilitation of co-circular line segments also takes place due to the angular dispersion of excitation off the orientation preference axis. The parameters $r_{1}$ and $r_{2}$ determine the minimum and maximum range of connections---$r_{1}$ is a short-range cutoff intended to eliminate self-excitation of short line segments, $\psi$ specifies the angular dispersion of excitation off the orientation preference axis, while $w_{e}>0$ and $w_{i}<0$ are the strengths of excitation and inhibition, respectively.
\begin{figure}
\includegraphics[width=140pt, bb=72 166 589 650, clip=true]{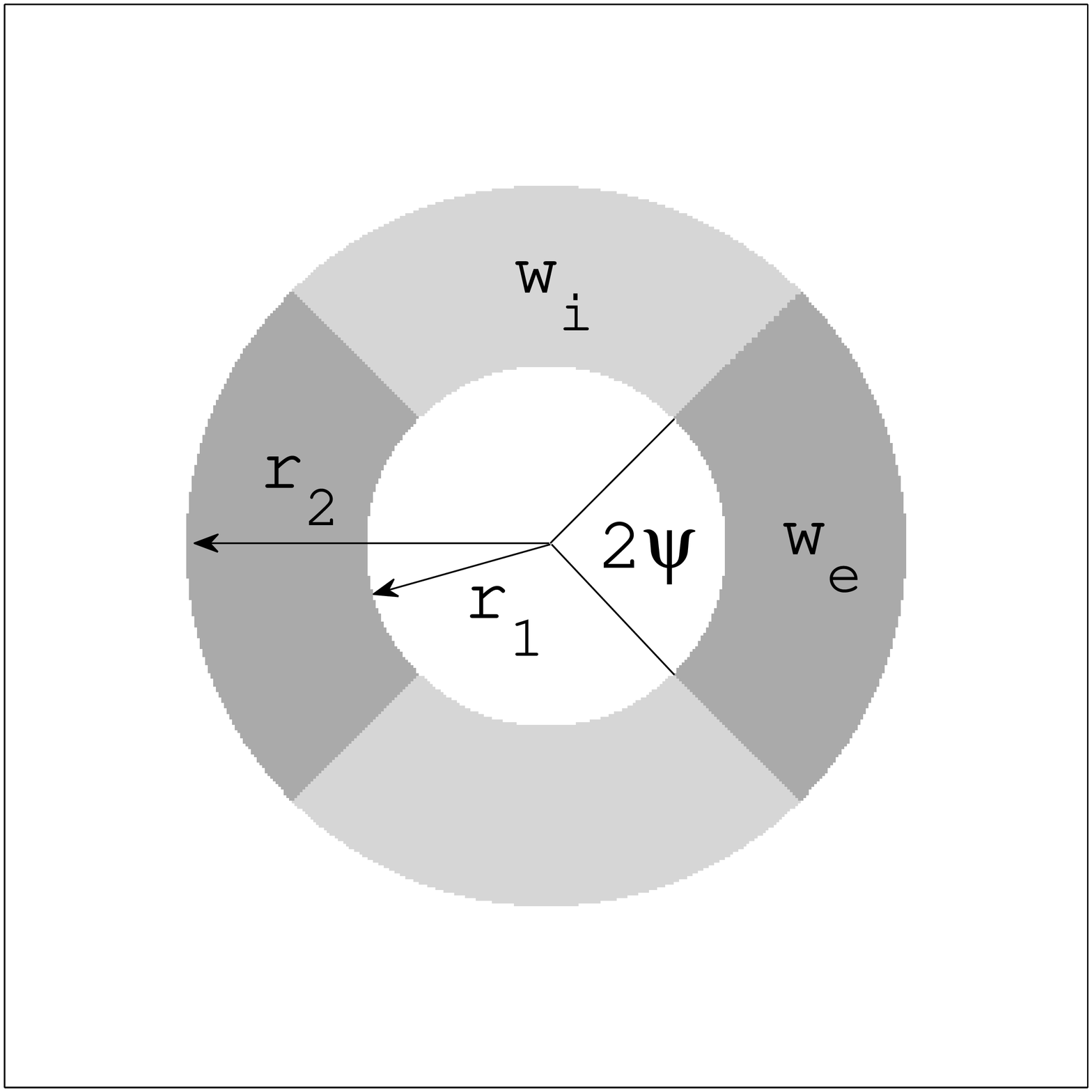}
\caption{Plot of the function $w({\bf{r}})$ used in Eq.~(\ref{connectivity}). The parameters $r_{1}$ and $r_{2}$ give the minimum and maximum range (arrows) of horizontal connections, $\psi$ is the angular dispersion of excitation off the orientation preference axis (shown as the horizontal axis), while $w_{e}$ and $w_{i}$ are the strengths of excitation and inhibition, respectively.}
\end{figure}

\subsection{Salient Line}\label{sec3.1}

Now we apply our model to group co-linear line segments which form a salient line. The image in Fig.~1(a) contains a single salient line made of 4 co-linear line segments within a cluttered background of 104 line segments with random orientations and positions. The image is normalized to have side lengths of 1 unit, setting the length scale for parameters describing the receptive field size and horizontal connection range. The size of each line segment is 0.0417$\times$0.0125, and the value of $I({\bf{r}})$ is 1 on a line segment, and 0 elsewhere.

To detect the salient line in Fig.~1(a) the first step is to find $I_{\phi}({\bf{r}})$---that is, for each value of $\phi$ we find the set of line segments which trigger the largest neural activity response. We discretize $\phi$ into $k$ elements between 0 and $\pi$: $\phi=0$, $\pi/k$, $2\pi/k$,..., $(k-1)\pi/k$; and set $k=10$. Applying Eqs.~(\ref{gabor})--(\ref{endgabor}) to the image in Fig.~1(a) yields $I_{\phi}({\bf{r}})$, as shown in Fig.~3(a) for all values of $\phi$. This corresponds to the simple-cell neural activity. Next, applying Eqs.~(\ref{eqlm})--(\ref{totalact}) yields the neural activity $u({\bf{r}})$ from the feed-forward model, as shown in Fig.~3(b). The model parameters are given in the figure caption, and the values of $\psi$ and $r_{1}$ were chosen to facilitate co-linearity and a short-range cutoff (where only line segments separated by distance $d\geq r_{1}$ are facilitated). Several groups of co-linear line segments above the cutoff range $r_{1}$ have been facilitated by horizontal connections in Fig.~3(b), but only one of these corresponds to the salient line in Fig.~1(a)---the rest are fortuitous alignments of background clutter segments. Finally, iterating Eqs.~(\ref{eqlm}) and (\ref{thresh}) 4 times yields $u({\bf{r}})$ from the feedback model, as shown in Fig.~3(c). Cooperative interactions between facilitated line segments take place over the 4 iterations until the only activity that remains corresponds to the salient line in Fig.~1(a).
\begin{figure}
\includegraphics[width=150pt,bb=250 244 370 618,clip=true]{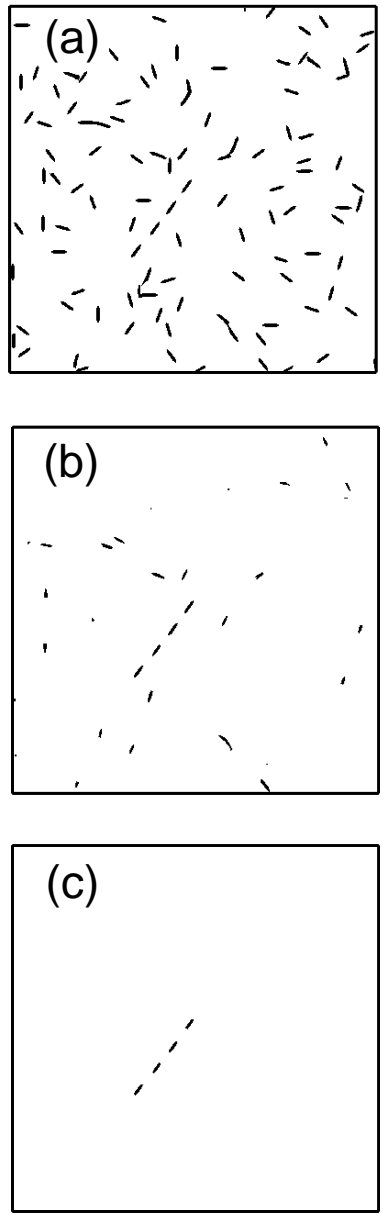}
\caption{Saliency assigned by different stages of the V1 model for the image in Fig.~1(a). (a) Simple cell activity $I_{\phi}({\bf{r}})$ for all $\phi$ from Eqs.~(\ref{gabor})--(\ref{endgabor}) with Fig.~1(a) for $I({\bf{r}})$. (b) Feed-forward activity $u({\bf{r}})$ from Eqs.~(\ref{eqlm})--(\ref{totalact}). Only highest activity is shown. (c) Feedback activity $u({\bf{r}})$ from iterating Eqs.~(\ref{eqlm}) and (\ref{thresh}) 4 times. Parameters are: $\kappa=0.06$, $\gamma=0.5$, $\sigma=0.008$, $\lambda=0.025$, $\varphi=0$, $r_{1}=0.125$, $r_{2}=0.175$, $\psi=\pi/12$, $w_{e}=0.5$, $w_{i}=0$, $\phi_{max}=\pi/12$, $\theta=0.6$, and $u=0.44$.}
\end{figure}

\subsection{Circular Closed Contour}\label{sec3.2}

The model is now used to group line segments which form a closed contour that is approximately circular. The image in Fig.~1(b) contains a salient closed contour made from 10 line segments within a cluttered background of 97 line segments with random orientations and positions. Image details are the same as in Sec.~\ref{sec3.1}.

Applying Eqs.~(\ref{gabor})--(\ref{endgabor}) to the image in Fig.~1(b) yields $I_{\phi}({\bf{r}})$, as shown in Fig.~4(a) for all values of $\phi$. Next, applying Eqs.~(\ref{eqlm})--(\ref{totalact}) yields the neural activity $u({\bf{r}})$ from the feed-forward model, as shown in Fig.~4(b). The model parameters are given in the figure caption, and the values of $\psi$ and $\phi_{max}$ were chosen to facilitate line segments which form circular arcs as well as lines. Several groups of line segments forming circular arcs and lines have been facilitated by horizontal connections in Fig.~4(b), however, only one of these corresponds to the closed contour in Fig.~1(b). As shown in the previous subsection, the feed-forward model filters out much of the background clutter from the image, while retaining the salient feature.
\begin{figure}
\includegraphics[width=150pt, bb=254 256 363 600,clip=true]{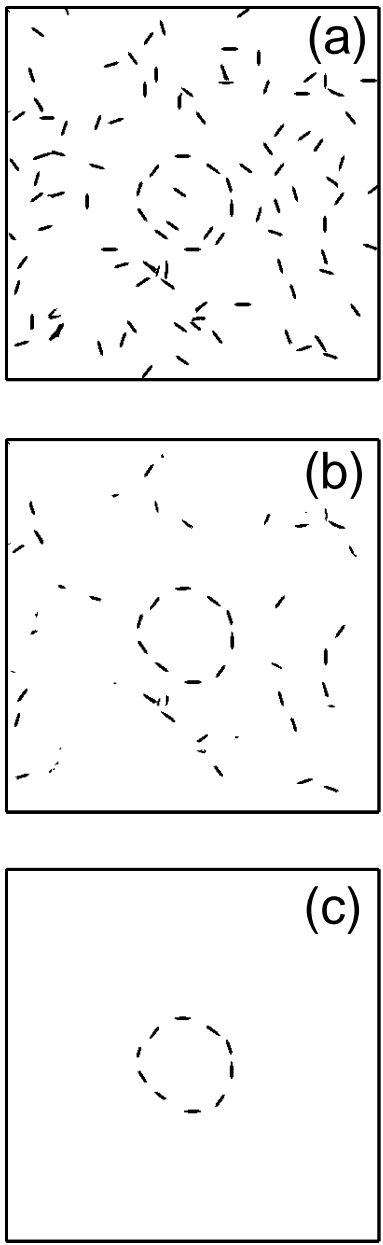}
\caption{Saliency assigned by different stages of the V1 model for the image in Fig.~1(b). (a) Simple cell activity using Fig.~1(b) for $I({\bf{r}})$. (b) Feed-forward activity from Eqs.~(\ref{eqlm})--(\ref{totalact}). Only highest activity is shown. Parameters are: $r_{1}=0.042$, $r_{2}=0.117$, $\psi=\pi/6$, $w_{i}=-0.3$, $\phi_{max}=\pi/5$, and others as in Fig.~3. (c) Feedback activity from iterating Eqs.~(\ref{eqlm}) and (\ref{thresh}) 7 times with simple cell activity from Fig.~5(a). Parameters are now: $r_{2}=0.133$, $\psi=\pi/4$, $\phi_{max}=\pi/3$, $u=0.473+0.017n$ (for iterate $n$) and others as in (b).}
\end{figure}

The feedback model activity is shown in Fig.~4(c), and results from iterating Eqs.~(\ref{eqlm}) and (\ref{thresh}) 7 times with the simple cell activity from Fig.~4(a) and a different set of parameter values. It is seen that the only activity remaining corresponds to the salient closed contour in Fig.~1(b). The model parameters are given in the figure caption, and the values of $r_{2}$, $\psi$, and $\phi_{max}$ have been increased relative to their values in Fig.~4(b). The reason for this was to increase the size of the excitatory component of the horizontal connections in order to increase facilitation between neighboring line segments in the closed contour. This can be seen in Fig.~5, where the bottom line segment of the closed contour is completely facilitated by its left and right neighbors. A general principal now applies: Each line segment that is part of a closed contour is facilitated by two neighbors, while line segments at the ends of an open contour are facilitated by only one neighbor. This principal allows cooperative interactions in the feedback model to group line segments according to whether or not they form closed contours. Open contours get shorter and shorter with each iteration of feedback as line segments are discarded from the ends. A closed contour will also become inhibited less and less as nearby clutter segments are iteratively discarded. In order to maintain a balance between excitation and inhibition in this model, the strength of inhibition in the thresholding step needs to be increased as inhibition due to horizontal connections gradually decreases. This is done linearly in the number of iterations using the form for $u$ given in Fig.~4.
\begin{figure}
\includegraphics[width=150pt,bb=250 293 366 537]{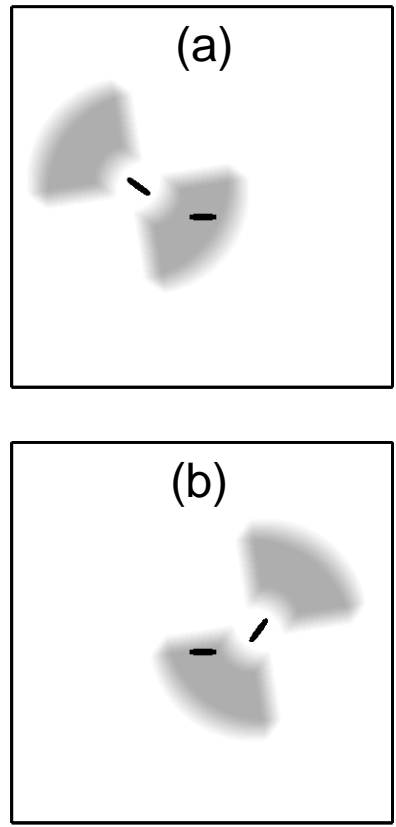}
\caption{Facilitation of the bottom line segment in Fig.4(c) by its left and right neighbors. (a) and (b) show an enlargement of the excitatory component of horizontal connections (Grey) due to each neighbor. Parameters are as in Fig.~4(c).}
\end{figure}

\subsection{Non-Circular Closed Contour}\label{sec3.3}

The model is now used to group line segments which form a closed contour that is not even approximately circular, but has an arbitrary shape. The image in Fig.~1(c) consists of several curved segments, some making up the closed contour and the rest forming background clutter. Applying Eqs.~(\ref{gabor})--(\ref{endgabor}) to the image in Fig.~1(c) yields $I_{\phi}({\bf{r}})$, as shown in Fig.~6(a) for all values of $\phi$. Next, applying Eqs.~(\ref{eqlm})--(\ref{totalact}) yields the neural activity $u({\bf{r}})$ from the feed-forward model, as shown in Fig.~6(b). The feedback model activity is shown in Fig.~6(c), and results from iterating Eqs.~(\ref{eqlm}) and (\ref{thresh}) 9 times. It is seen that this activity approximates the salient closed contour in Fig.~1(c).

The model parameters are given in the figure caption, and the value of $\phi_{max}$ indicates that horizontal connections are not orientation specific---line segments of all orientations are connected together. The reason is that the closed contour is not even close to being smooth or circular, and has several regions where line segments actually meet at right angles instead of having smooth circular joins. Models which are limited to co-circular connections would not be able to detect a closed contour in this case.

\begin{figure}
\includegraphics[width=150pt, bb=246 236 371 635,clip=true]{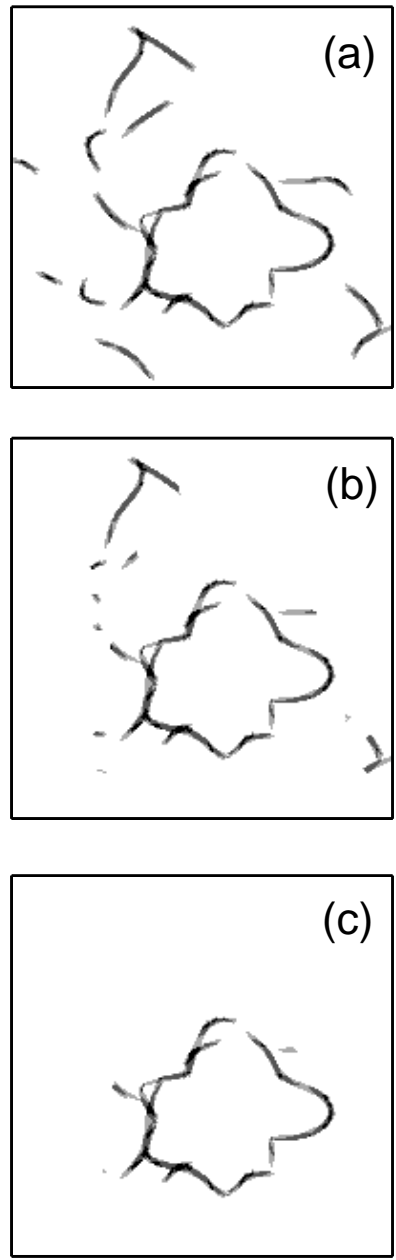}
\caption{(a) Simple cell activity using Fig.~1(c) for $I({\bf{r}})$. (b) Feed-forward activity from Eqs.~(\ref{eqlm})--(\ref{totalact}). Only highest activity is shown. (c) Feedback activity from iterating Eqs.~(\ref{eqlm}) and (\ref{thresh}) 9 times. Parameters are: $k=8$, $\sigma=0.017$, $\lambda=0.048$, $r_{1}=0.05$, $r_{2}=0.117$, $\psi=\pi/3$, $w_{i}=0$, $\phi_{max}=\pi/2$, $u=0.66$, and others as in Fig.~3.}
\end{figure}

\section{Summary and Discussion}\label{sec4}

In this work, our primary aim has been to establish a solid foundation for modeling horizontal connections in biological neural networks. Our model takes into account the known classical receptive fields of simple cells in V1, the activity dynamics of neural populations, and some of the basic properties of horizontal connections found in physiological and psychophysical studies. 

We have applied our model to the detection of salient contours in three different cases: a line, a circular closed contour, and a non-circular closed contour. Using different values of the model parameters the model performed well in all cases: assigning higher neural activity to line segments forming a salient contour, and lower activity to line segments forming background clutter. Values of the model parameters for each case indicated different facilitation mechanisms taking place. Horizontal connections facilitating co-linear line segments and a short-range cutoff were useful for detecting salient lines. Facilitation of line segments forming circular arcs was found to be useful for detecting salient closed contours which were approximately circular. In addition, cooperative interactions were able to distinguish between open and closed contours---open contours getting shorter and shorter with each iteration as line segments were discarded from the ends, while closed contours remained unchanged. For the detection of non-circular closed contours facilitation of line segments of all orientations was found to be necessary. The reason was that there was no guarantee of having smooth circular joins between different line segments in this contour.

Finally, we showed how our model can operate as either a feed-forward or a feedback network. Operating as a feed-forward network, the model assigns saliency within 10 ms, and increases the neural activity of line segments facilitated by the horizontal connections. In Figs.~3(b) and 4(b), the feed-forward network filtered out much of the background clutter while retaining the salient contour. Operating as a feedback network, further improvement is achieved through cooperative interactions over several iterations. Further, our model does not suffer from stability problems as dynamical neural models sometimes do. This makes it suitable for use in assigning saliency in biologically-inspired neural networks performing object recognition.

Several further questions remain. Firstly, it is difficult to know what model parameter values are actually used in computations involving V1 horizontal connections. It seems unlikely that precise definitions such as co-circularity are implemented by horizontal connections---this property would only apply to a small number of all salient contours. We favor a less stringent construction involving a modified association field. Also, scale invariance has been observed in natural images. To include this in our model would require a range of receptive field sizes and horizontal connection ranges to be considered.

The performance of our model in the successful detection of a salient feature depends on details of the background clutter relative to the feature. For example, operating as a feedback network, the number of iterations taken to distinguish between open and closed contours will depend directly on contour length. A long open contour will take a larger number of iterations to resolve than a short one. Similarly, detecting the most salient feature from a set of salient features could also take a large number of iterations. Addressing these questions, and using bottom-up saliency to robustly improve performance in object recognition tasks, will be the subject of future work.

\begin{acknowledgments}
We would like to thank S.~Barr for the image in Fig.~1(c), and G.~Kenyon and S.~Brumby for useful related discussions. We gratefully acknowledge the support of the U.S. Department of Energy through
the LANL/LDRD Program project 20090006DR for this work. 
\end{acknowledgments}

\end{document}